\documentclass[prl,superscriptaddress,twocolumn,amsmath,amssymb]{revtex4}
\usepackage{amsthm}
\usepackage{graphicx}

\newcommand{\bra}[1]{\langle #1 |}
\newcommand{\ket}[1]{| #1 \rangle}

\newcommand{\tr}{\text{Tr}}
\newcommand{\one}{{\bf 1}}

\newcommand{\prlsection}[1]{{\it{#1}} ---}

\newcommand{\Hil}{\mathcal H}
\newcommand{\channel}{\mathcal E}
\newcommand{\channelb}{\mathcal E^\dagger}

\newcommand{\cref}[1]{(\ref{#1})}

\newcommand{\zeroset}[1]{{\mathbb K}}

\newtheorem{thm}{Theorem}

\begin{document}

\title{Generalization of Quantum Error Correction via the Heisenberg Picture}
\author{C\'edric B\'eny}
\affiliation{Department of Applied Mathematics, University of
Waterloo, ON, Canada, N2L 3G1}
\author{Achim Kempf}
\affiliation{Department of Applied Mathematics, University of
Waterloo, ON, Canada, N2L 3G1}
\author{David~W.~Kribs}
\affiliation{Department of Mathematics and Statistics, University
of Guelph, Guelph, ON, Canada, N1G 2W1} \affiliation{Institute for
Quantum Computing, University of Waterloo, ON Canada, N2L 3G1}
\date{\today}

\begin{abstract}

We show that the theory of operator quantum error correction can
be naturally generalized by allowing constraints not only on
states but also on observables. The resulting theory describes the
correction of algebras of observables (and may therefore suitably
be called ``operator algebra quantum error correction''). In
particular, the approach provides a framework for the correction
of hybrid quantum-classical information and it does not require
the state to be entirely in one of the corresponding subspaces or
subsystems. We discuss applications to quantum teleportation and
to the study of information flows in quantum interactions.

\end{abstract}

\maketitle

Error correction methods are of crucial importance for quantum
computing and the so far most general framework, called operator
quantum error correction (OQEC) \cite{kribs05, kribs06},
encompasses
active error correction \cite{bennett96,knill97,shor95, steane96,
gottesman96} (QEC), together with the concepts of decoherence-free
and noiseless subspaces and subsystems \cite{zanardi97, palma96,
duan97, lidar98,knill00, zanardi01a, kempe01}. The OQEC approach
has enabled more efficient correction procedures in active error
correction \cite{bacon06,poulin05x1,nielsen05,andreas06}, has led
to improved threshold results in fault tolerant quantum computing
\cite{AC06}, and has motivated the development of a structure
theory for passive error correction \cite{choi05, knill06} which
has recently been used in quantum gravity \cite{dreyer06,
poulin05b, kribs05x1, konopka06}.

In this paper, we introduce a natural generalization of this theory.  To this end, we change the focus from that of states to that of observables: conservation of a state by a given noise model implies the conservation of all of its observables, and is therefore a rather strong requirement. This can be alleviated by specifically selecting only some observables to be conserved. 
In this context it is  natural to consider {\it algebras} of observables \cite{vN55}. 
Hence our codes take the form of operator algebras that are closed under Hermitian conjugation; that is, finite dimensional C$^*$-algebras \cite{Dav96}. As a convenience, we shall simply refer to such operator algebras as ``algebras''. 
Correspondingly we refer to the new theory as ``operator algebra quantum error correction'' (OAQEC). 
We present results that establish testable conditions for
correctability in OAQEC. We also discuss illustrative examples and
consider applications to quantum teleportation and information
flow in quantum interactions. We shall present the proofs and more
examples in \cite{BKK07b}.


Noise models in quantum information are described by {\it channels}, which are (in the Schr\"odinger picture) trace-preserving (TP) and completely positive (CP) linear maps $\channel$ on mixed states, $\rho$, which are operators acting on a Hilbert space $\Hil$. 
If $\rho$ is a density matrix we can always write $\rho \mapsto
\channel(\rho) = \sum_a E_a \rho E_a^\dagger$ where $\{E_a\}$ is a
non-unique family of channel {\em elements}. The QEC framework
addresses the question of whether a given subspace of states
$P\Hil$, called the code, can be corrected in the sense that there
exists a correction channel $\mathcal R$ such that $\mathcal R
(\channel (\rho)) = \rho$ for all states $\rho$ in the subspace;
that is, all $\rho$ which satisfy $\rho = P\rho P$. This amounts
to asking for a subspace on which $\channel$ has a left inverse
that is a physical map. From QEC to OQEC the scope of error
correction is generalized by only requiring the states of a
subsystem to be conserved: $\mathcal R (\channel (\rho \otimes
\tau)) = \rho \otimes \tau'$ for all $\rho \otimes \tau$ in the
subspace.
As we will show, this amounts to the correction of special types of algebras. 
In general, every algebra $\mathcal A$ of observables induces a
decomposition of the Hilbert space $\mathcal H$ into ${\mathcal H}
= \oplus_{k=1}^d (A_k\otimes B_k) \oplus C$. Here all operators in
the algebra have $C$ in their kernel and act irreducibly on each
$A_k$ while acting trivially on the subsystem $B_k$. This means
that the algebra decomposes as
\begin{equation}\label{algebraform}
\mathcal A = \oplus_{k=1}^d(\mathcal L(A_k)\otimes\one^{B_k})
\oplus 0_C,
\end{equation}
where  $\mathcal L(A_k)$ denotes the set of all operators on
$A_k$, $\one^{B_k}$ is the identity operator on $B_k$, and $0_C$
is the zero operator on $C$. From this perspective, we can view
the QEC framework as focussing on codes $\mathcal
L(A)\otimes\one^B$ (or subspaces $A$) with $\dim B=1$. Moreover,
OQEC considers ``subsystem codes'' encoded in algebras of the form
$\mathcal L(A)\otimes\one^B$ for general subsystems $A$ and $B$.
Classical information is captured by the case in which $\dim A
=1$: commutative algebras. Thus, in addition to the classical,
QEC, and OQEC cases, our new OAQEC approach also provides a
framework for the correction of hybrid quantum-classical
information and memory \cite{Kuper03} exposed to external noise.
In particular, this includes cases in which separate (orthogonal)
parcels of quantum information are labelled by classical
``addresses''.

Let us discuss in more detail the motivation for considering algebras. 
We begin by recalling that a general observable is a positive
operator-valued measure (POVM) $X(\Delta)$ where $\Delta \subset
\Omega$, the set in which the observable $X$ takes values. For
simplicity we consider observables with a finite number of
outcomes which can be characterized by a family of positive
operators $\{X_a\}$. In the Heisenberg picture an observable
evolves according to the unital CP-map $\channelb$ with elements
$E_a^\dagger$ instead of $E_a$. If for all values of the label $a$
there exists an operator $Y_a$ such that $X_a = \channelb(Y_a)$
then all the statistical information about $X$ has been conserved
by $\mathcal E$ since for any initial state $\rho$ we have,
$\tr(\rho X_a) = \tr(\rho\channel^\dagger(Y_a)) =
\tr(\channel(\rho)Y_a)$, the latter equality following from the
definition of ${\mathcal E}^\dagger$. In this case, to correct for
the errors induced by $\mathcal E$ we need a channel $\mathcal R$
that maps each $X_a$ to one of the operators $Y_a$ through
$\mathcal R^\dagger(X_a) = Y_a$, so that $(\mathcal R \circ
\channel)^\dagger(X_a) =({\mathcal E}^\dagger\circ{\mathcal
R}^\dagger)(X_a) = X_a$. In such a scenario, we will say that
$X_a$ is {\em correctable} for $\channel$ and {\it conserved} by
$\mathcal R \circ \channel$. In particular, if $X$ is a standard
projective measurement, and so $X_a^2 = X_a$ for all $a$, then the
projectors $X_a$ linearly span the algebra they generate. Hence,
in this case $\mathcal R \circ \channel$ conserves an entire
commutative algebra. Therefore, focussing on the correctability of
sets of observables which have the structure of an algebra,
apart from allowing a complete characterization, is also
sufficient for the study of all the correctable {\em projective}
observables.

One result of this paper will be to show that there always exists
a single channel $\mathcal R$ correcting all projective
observables correctable in the above sense. In fact, we study a
more general question: If we have some control on the initial
states, which is expected in a quantum computation, then we can
ask for an observable to be conserved only if the state starts in
a certain subspace $P\Hil$. That is, $P(\channelb \circ \mathcal
R^\dagger)(X)P = PXP$. We derive a necessary and sufficient
condition for an entire algebra of operators on $P\Hil$ to be
simultaneously correctable in that sense. The resulting theory
contains OQEC and QEC in the special cases discussed above.

We remark that our approach differs from that of the stabilizer
formalism \cite{gottesman96} where observables in the Heisenberg
picture are used as a way to characterize a subspace of states.
Our approach is closer in spirit to that of~\cite{deutsch99}. The
idea that observables naturally characterize subsystems has also
been exploited in~\cite{zanardi04x1}.

\prlsection{Noiseless subsystems} First we recall the definition
of a noiseless subsystem and we give an equivalent definition in
terms of the dual channel $\channelb$. Consider a decomposition of
a finite-dimensional Hilbert space $\Hil$ as $\Hil = (A \otimes B)
\oplus C$. We introduce the projector $P$ on the subspace $A
\otimes B$.
By definition, $A$ is a noiseless subsystem for $\channel$ if for
all $\rho \in {\mathcal L}(A)$ and $\sigma \in {\mathcal L}(B)$
there exists $\tau \in {\mathcal L}(B)$ such that $\channel(\rho
\otimes \sigma) = \rho \otimes \tau$. In terms of the dual channel
$\channelb$, the subsystem $A$ is noiseless for $\channel$ if and
only if
\begin{equation}\label{oqecns}
P\, \channelb(X \otimes \one)\, P = X \otimes \one
\end{equation}
for all operators $X$ acting on $A$. This is a consequence of the
noiseless subsystem characterization from \cite{kribs05} as can be
readily verified.
In Eq.~(\ref{oqecns}), the projectors $P$ are needed since the
definition of the noiseless subsystem is only concerned with what
happens to states initially in the subspace $A \otimes B$. In
general, an initial component outside this space may, after
evolution, disturb the otherwise noiseless observables.

\prlsection{Conserved observables} If $P\channelb(X_a)P = PX_aP$
for all $a$ we say that the observable $X$ is {\em conserved} by
$\channel$ for states in $P\Hil$. More generally, let us say an
algebra $\mathcal A$ is {\em conserved} by $\channel$ for states
in $P\Hil$ if every element of $\mathcal A$ is conserved; that is,
\begin{equation}\label{conservedalg}
P\,\channelb(X)\,P = PXP \quad \forall X \in \mathcal A.
\end{equation}

Notice that Eq.~(\ref{conservedalg}) gives a generalization of
noiseless subsystems. Indeed, any subalgebra $\mathcal A$ of
$\mathcal L(P\Hil)$ for which all elements $X\in \mathcal A$
satisfy Eq.~(\ref{conservedalg}) is a direct sum of noiseless
subsystems. This can be seen by first noting that any algebra
$\mathcal A$ has a decomposition of the form given in
Eq.~(\ref{algebraform}), and then applying Eq.~(\ref{oqecns}). In
particular, focussing on the so-called ``simple'' algebras
$\mathcal L(A) \otimes \one^B$ captures standard noiseless
subsystems as in Eq.~(\ref{oqecns}).

The following theorem provides testable conditions that
characterize when an algebra is conserved on states in a given
subspace by a channel, strictly in terms of the operation elements
for the channel. The result comes as an adaptation of results from
\cite{choi05} and we shall present its proof in \cite{BKK07b}. It
is a generalization because, here, the algebra need not contain
the projector $P$.

\begin{thm}
\label{cons} A subalgebra $\mathcal A$ of $\mathcal L(P\Hil)$ is
conserved on states in $P\Hil$ by a channel $\mathcal E$ if and
only if $[E_a P, X] = 0$ for all elements $E_a$ and all $X \in
\mathcal A$.
\end{thm}

Heuristically, an algebra supported on a subspace is conserved by
a channel precisely when elements of the algebra commute with the
generators of the noise, restricted to the subspace.

\prlsection{Error correction of observables} We say that an
algebra $\mathcal A$ is {\em correctable} for $\mathcal E$ on
states in the subspace $P\Hil$  if there exists a channel
$\mathcal R$ such that
\begin{equation}\label{Heiseneqn}
P(\mathcal R \circ \channel)^\dagger(X)P = PXP \quad \forall
X\in\mathcal A.
\end{equation}
This notion of correctability is more general than the one
addressed by the framework of OQEC. Indeed, OQEC focusses on
simple algebras, $\mathcal L(A)\otimes\one^B$. Here,
correctability is defined for any finite-dimensional algebra. A
further generalization is that we do not require $P$ to belong to
the algebra considered. We expand on these points in \cite{BKK07b}
via an examination of the Schr\"odinger formulation of OAQEC.

We now state the main result of the paper, which generalizes the
fundamental result for both QEC \cite{knill97} and OQEC
\cite{kribs05, nielsen05}. It provides conditions for testing
whether an algebra is correctable for a given channel in terms of
its operation elements.

\begin{thm}
\label{found} A subalgebra $\mathcal A$ of $\mathcal B(P\Hil)$ is
correctable on $P\Hil$ for the channel $\channel$ if and only if
\begin{equation}
\label{found:equ} [P E_c^\dagger E_b^{} P, X] = 0 \quad \forall
X\in{\mathcal A}\,\,\,\, \forall c,b.
\end{equation}
\end{thm}

We present the proof in \cite{BKK07b}. Not surprisingly, the
operation elements $\{E_c^\dagger E_b\}$ for ${\mathcal
E}^\dagger\circ{\mathcal E}$ play a key role as in other error
correction settings.
We also note that the correction channel $\mathcal R$ constructed in the proof corrects any channel whose elements are linear combinations of the elements $E_a$. Thus, as in the original theory of error correction, we can in practice neglect the channel $\channel$ and focus instead directly on the discrete {\em error operators} $E_a$.  

It is instructive to consider the special case of classical OAQEC
codes. A classical channel has elements $E_{ij} = \sqrt{p_{ij}}
\ket{i}\bra{j}$, where $(p_{ij})$ forms a stochastic matrix with
transitional probabilities $p_{ij}$ from $j$ to $i$. Thus,
$E_{ij}^\dagger E_{kl} = \delta_{ik} \sqrt{p_{ij}p_{il}}
\ket{j}\bra{l}$, and Theorem~\ref{found} shows that if $\alpha =
(\alpha_j)$ are the diagonal components of a classical (diagonal)
observable, then $\alpha$ can be corrected if and only if
$\alpha_j = \alpha_k$ for all $k,j$ such that there is an $i$ with
$p_{ij}\neq 0$ and $p_{ik}\neq 0$. Heuristically, two states
cannot be distinguished from each other after the channel has
acted precisely when there is a nonzero probability of a
transition from both states to a common state.

Does OAQEC offer more powerful error correction procedures than
OQEC? It is easy to see that if an algebra $\mathcal A$ is
correctable according to this scheme then each simple sector
$\mathcal L(A)\otimes\one^B$ is individually correctable through
OQEC (or QEC when $\dim B=1$). However, OAQEC codes have at least
two attractive features: First, all simple sectors can be
corrected simultaneously by the {\it same} correction channel.
Secondly, each simple sector can be corrected even if the initial
state is not entirely in the corresponding subspace. In
particular, the initial state could be in a quantum superposition
between various sectors even though combined sectors may not be
correctable in the traditional sense.

As an illustrative example, consider a 2-qubit system exposed to noise inducing a phase flip error $Z$ (a unitary Pauli operator) on the first qubit with nonzero probability $p$. 
The noise model $\mathcal E$ has elements $\{\sqrt{1-p}\,\,\one,
\sqrt{p} \,\,Z_1 \}$. Let ${\mathcal C}_1$ be the (subspace) code
with basis $\ket{0_L}=\ket{00}$, $\ket{1_L}=\ket{01}$ and
${\mathcal C}_2$ the code with basis $\ket{0_L}=\ket{10}$,
$\ket{1_L}=\ket{11}$. It is clear that each of ${\mathcal C}_1$
and ${\mathcal C}_2$, or from the OAQEC perspective the algebras
${\mathcal L}({\mathcal C}_1)$ and ${\mathcal L}({\mathcal C}_2)$,
is correctable individually for $\mathcal E$. Indeed, each code is
a stabilizer subspace for $Z_1$ (for eigenvalue $1$ and $-1$
respectively). However, the combined code ${\mathcal
C}_1\oplus{\mathcal C}_2$, or equivalently ${\mathcal L}({\mathcal
C}_1\oplus{\mathcal C}_2)={\mathcal L}(\mathbb C^4)$, is clearly
not correctable for $\mathcal E$ because $\mathcal E$ is not a
unitary operation.

Nevertheless, the hybrid qubit-qubit OAQEC code ${\mathcal A} =
{\mathcal L}({\mathcal C}_1)\oplus {\mathcal L}({\mathcal C}_2)$
{\it is} correctable for $\mathcal E$. This immediately follows
from Theorem~\ref{found} and the observation that $Z_1$ belongs to
the set of operators which commute with $\mathcal A$: the
commutant ${\mathcal A}^\prime =  {\mathbb C} P_1 \oplus {\mathbb
C} P_2$, where we have written $P_i$ for the projector onto
${\mathcal C}_i$. As discussed above, separate (orthogonal) codes
that determine the simple summands of an OAQEC code can all be
corrected by the same correction operation, which is clearly not
the case in general for non-OAQEC codes. Moreover, as in QEC and
OQEC, the theory includes an explicit recipe for constructing the
correction operation. This point will be elucidated further in
\cite{BKK07b}. In this case one can easily verify that the channel
${\mathcal R}$ with elements $\{P_1,Z_1P_2\}$ satisfies
$({\mathcal R}\circ{\mathcal E})^\dagger (X) = X$ for all
$X\in\mathcal A$.

\prlsection{Application: noisy quantum teleportation} In the
standard picture for quantum teleportation \cite{BBC93}, Alice can
send Bob an entire qubit by sending two classical bits, provided
they share a maximally entangled pair of qubits initially.
Consider a pair of qubits in the maximally entangled state
$\ket{\psi} = \frac{1}{\sqrt{2}}(\ket{00} + \ket{11})$. We assume
that Alice and Bob each possess one qubit of this pair. Consider
also the  unitary Pauli operators $\{U_0=\one, U_1=X, U_2=Y,
U_3=Z\}$.
In the language of channels, teleportation can be viewed as the channel $\mathcal E$ that has for input the qubit $\ket{\psi}$ to be teleported and for output three qubits: two which represent the bits Alice needs to send to Bob, and Bob's qubit: 
$ \mathcal E(\ket{\psi}\bra{\psi}) = \frac{1}{4} \sum_{i=0}^3
\ket{i}\bra{i} \otimes U_i (\ket{\psi}\bra{\psi}) U_i^\dagger,
$
where $\ket{i}$ form an orthonormal basis of the four-dimensional
Hilbert space representing the two bits that Alice must send to
Bob over a classical channel. We can readily verify that Bob can
indeed fully correct the channel. The channel elements are of the
form $E_i \propto \ket{i}\otimes U_i$, and hence the condition of
Theorem~\ref{found} is met since
$ E_i^\dagger E_j \propto \bra{i}\ket{j} \otimes U_i^\dagger U_j =
\delta_{ij} U_i^\dagger U_i = \delta_{ij}\one.
$ 
Thus all operators on the initial qubit can be corrected, and all
the information can be recovered. Note that even though the
operators $E_i$ map between two different spaces, the operators
$E_i^\dagger E_j$ map the initial qubit space to itself. One can
teleport as many qubits as desired in parallel provided that one
starts with one shared entangled pair per qubit to be teleported.

Consider now the case in which the classical step of the
teleportation process, when Alice transmits bits to Bob, is
implemented over a noisy classical channel. We shall also assume
unitary encodings are taken from more general families of unitary
operators. To this end, consider a family $\{U_g\}_{g\in S}$ of
unitary operators in $\mathcal L(\mathcal H)$ for some Hilbert
space $\mathcal H$.
In the standard teleportation protocol, $S$ is the set of single
qubit Pauli operators. Now suppose that a noisy stochastic channel
is applied on the classical bits $\ket{g}$, $g\in S$, that Alice
must send to Bob with transition probabilities $p_{gh}$. Then one
could ask, what information can be recovered by Bob once he is in
possession of all the data?
The overall channel is given by composing $\mathcal E$ with the
classical channel. Thus, the error operators $F_{ghl}$ satisfy
$ F_{ghl} \propto \sqrt{p_{gh}} \ket{g}\bra{h}\ket{l}\otimes U_l =
\delta_{hl} \sqrt{p_{gh}} \ket{g} \otimes U_h.
$  
It follows that the commutant of the correctable algebra, which is
contained in $\mathcal L(\mathcal H)$, has generators
$ F^\dagger_{ghl} F_{g'h'l'} = \delta_{gg'} \sqrt{p_{gh}p_{gh'}}
U_h^\dagger U_{h'}.
$ 
We are only interested in the nonzero generators; that is, those
for which $p_{gh}\neq 0$ and $p_{gh'}\neq 0$. This condition means
that classically, because of the noise, we can no longer
distinguish between the state $\ket{h}$ and the state $\ket{h'}$.
Therefore, the commutant of the conserved algebra has a generator
$U_g^\dagger U_h$ for each pair of classical states $g,h$ that
became indistinguishable under the noisy channel. In short, if Bob
is not sure whether Alice's classical message was $g$ or $h$, then
he can only completely recover those properties of the quantum
states that are invariant under the transformations $U_g^\dagger
U_h$. In particular, this implies more general code algebras will
be obtained. We shall further investigate this ``noisy'' version
of quantum teleportation in \cite{BKK07b}.

\prlsection{Application: information flow in interactions}
Consider the interaction $U$ between a system $S$ and an apparatus
$A$ where the initial state of the apparatus is known to be
$\rho_A$. Tracing out either over $A$ or $S$ after the evolution
yields respectively the channel $\channel_{SS}(\rho_S) = \tr_A( U(
\rho_S \otimes \rho_A) U^\dagger)$ from $S$ to $S$ or
$\channel_{SA}(\rho_S) = \tr_S( U( \rho_S \otimes \rho_A)
U^\dagger)$ from $S$ to $A$.




Using OAQEC we can determine what observables of the system can be
corrected for either of the two channels.
The algebra $\mathcal A_{SA}$ preserved by $\channel_{SA}$, that
represents the information about $S$ which is transferred to $A$,
can be computed to be the largest algebra of operators commuting
with the range of $\channel_{SS}^\dagger$.
Hence a direct consequence of Theorem~\ref{found} is that in an
open dynamics defined by a channel $\channel$, full information
about a projective observable can escape the system if and only if
it commutes with all the operators in the range of the channel;
that is, those observables whose first moment is correctable for
$\channel$.
This is a generalization of work in \cite{lindblad99}. Furthermore
this method characterizes those observables which are effectively
duplicated; in other words, those whose information stayed in $S$
and also flowed to $A$. They form the commutative algebra
$\mathcal A_{SS} \cap \mathcal A_{SA}$. Those observables have
been non-destructively measured by the system $A$. This analysis
has implications for the theory of decoherence \cite{giulini96,
zurek03}: a unique commutative algebra of observables emerges
naturally as characterizing the information which is shared
between the system and the environment after an interaction. This
suggests that the {\em pointer observables} should be defined not
just by their property of being stably encoded in the system but
also by the requirement that the information they represent is
transmitted to the environment. In this sense there is no {\em
basis ambiguity} \cite{zurek81} for the interpretation of a
unitary interaction as a measurement of the system by the
apparatus.

\prlsection{Outlook} We have presented a generalization of the
theory of operator quantum error correction that allows for the
correction of an arbitrary algebra of operators. Our main result
gives a characterization of correctable codes in this scheme.
Proofs and more applications will be provided in \cite{BKK07b}.
The recent experience with operator quantum error correction
suggests a reinvestigation of codes that have appeared in the
literature for possibly improved efficiency or other applications
enabled by this approach. We also suggest that the applications to
quantum teleportation and information flow presented here warrant
further investigation.
Furthermore, we have here focussed on algebras of operators rather
than general operator subspaces. It should be most interesting to
consider the possible conservation of the statistics of general
POVMs.

\prlsection{Acknowledgments} We thank our colleagues at IQC, PI,
UG, and UW for helpful discussions. This work was partially
supported by NSERC, PREA, ERA, CFI and OIT.


\begin{thebibliography}{10}



\bibitem{kribs05}
D. Kribs, R. Laflamme, and D. Poulin, Phys. Rev. Lett. {\bf 94},
180501
  (2005).

\bibitem{kribs06}
D.~W. Kribs, R. Laflamme, D. Poulin, and M. Lesosky, Quant. Inf.
\& Comp. {\bf
  6},  382  (2006).

\bibitem{VKL01} L. Viola, E. Knill, and R. Laflamme,
J. Phys. A \textbf{34}, 7067 (2001).

\bibitem{bennett96}
C.~H. Bennett, D.~P. DiVincenzo, J.~A. Smolin, and W.~K. Wootters,
Phys. Rev. A
  {\bf 54},  3824  (1996).

\bibitem{knill97}
E. Knill and R. Laflamme, Phys. Rev. A {\bf 55},  900  (1997).

\bibitem{shor95}
P.~W. Shor, Phys. Rev. A {\bf 52},  R2493  (1995).

\bibitem{steane96}
A.~M. Steane, Phys. Rev. Lett. {\bf 77},  793  (1996).

\bibitem{gottesman96}
D. Gottesman, Phys. Rev. A {\bf 54},  1862  (1996).

\bibitem{zanardi97}
P. Zanardi and M. Rasetti, Phys. Rev. Lett. {\bf 79},  3306
(1997).

\bibitem{palma96}
G. Palma, K.-A. Suominen, and A. Ekert, Proc. Royal Soc. A {\bf
452},  567
  (1996).

\bibitem{duan97}
L.-M. Duan and G.-C. Guo, Phys. Rev. Lett. {\bf 79},  1953 (1997).

\bibitem{lidar98}
D.~A. Lidar, I.~L. Chuang, and K.~B. Whaley, Phys. Rev. Lett. {\bf
81},  2594  (1998).

\bibitem{knill00}
E. Knill, R. Laflamme, and L. Viola, Phys. Rev. Lett. {\bf 84},
2525  (2000).

\bibitem{zanardi01a}
P. Zanardi, Phys. Rev. A {\bf 63},  12301  (2001).

\bibitem{kempe01}
J. Kempe, D. Bacon, D.~A. Lidar, and K.~B. Whaley, Phys. Rev. A
{\bf 63},
  42307  (2001).

\bibitem{bacon06}
D. Bacon, Phys. Rev. A {\bf 73}, 012340  (2006).

\bibitem{poulin05x1}
D. Poulin, Phys. Rev. Lett. {\bf 95}, 230504  (2005).

\bibitem{nielsen05}
M.~A. Nielsen and D. Poulin, e-print quant-ph/0506069  (2005).

\bibitem{andreas06}
A. Klappenecker and P.~K. Sarvepalli, e-print quant-ph/0604161
(2006).

\bibitem{AC06}
P. Aliferis and A.~W. Cross, e-print quant-ph/0610063 (2006).

\bibitem{choi05}
M.-D. Choi and D.~W. Kribs, Phys. Rev. Lett. {\bf 96},  050501
(2006).

\bibitem{knill06}
E. Knill, e-print quant-ph/0603252  (2006).

\bibitem{dreyer06}
O. Dreyer, F. Markopoulou, and L. Smolin, Nucl.Phys. B {\bf 744},
1  (2006).

\bibitem{poulin05b}
D. Poulin, e-print quant-ph/0505081  (2005).

\bibitem{kribs05x1}
D.~W. Kribs and F. Markopoulou, e-print gr-qc/0510052  (2005).

\bibitem{konopka06}
T. Konopka and F. Markopoulou, e-print gr-qc/0601028  (2006).

\bibitem{vN55}
J. von Neumann, {\it Mathematical foundations of quantum
mechanics}, Princeton University Press, Princeton, 1955.

\bibitem{Dav96}
K.~R. Davidson, {\it $C^*$-algebras by example}, Fields Institute
Monographs, 6. American Mathematical Society, Providence, RI,
1996.

\bibitem{BKK07b}
C. Beny, A. Kempf, and D.~W. Kribs, in preparation (2006).

\bibitem{Kuper03}
G. Kuperberg, IEEE Trans. Inform. Theory {\bf 49}, 1465 (2003).

\bibitem{deutsch99}
D. Deutsch and P. Hayden, Proc. R. Soc. Lond. A {\bf 456},  1759
(1999).

\bibitem{zanardi04x1}
P. Zanardi, D.~A. Lidar, and S. Lloyd, Phys. Rev. Lett. {\bf 92},
060402
  (2004).

\bibitem{BBC93}
C.~H. Bennett, G. Brassard, C. Crepeau, R. Jozsa, A. Peres, W.
Wootters, Phys. Rev. Lett. {\bf 70}, 1895 (1993).

\bibitem{lindblad99}
G. Lindblad, Lett. Math. Phys. {\bf 47},  189  (1999).


\bibitem{giulini96}
D. Giulini {\it et~al.}, {\em {Decoherence and the Appearance of a
Classical
  World in Quantum Theory}} (Springer, Berlin, 1996).

\bibitem{zurek03}
W.~H. Zurek, Rev. Mod. Phys. {\bf 75},  715  (2003).

\bibitem{zurek81}
W.~H. Zurek, Phys. Rev. D {\bf 24},  1516  (1981).

\end{thebibliography}

\end{document}